\documentclass{article}
\usepackage[psamsfonts]{amssymb}
\usepackage{amsmath}
\usepackage{cite}

\def\id{\mathbf{1}}

\def\ir{\mathrm{i}}

\begin{document}
\begin{titlepage}
\noindent{\large\textbf{TeV-photon paradox and space with SU(2) fuzziness}}

\vskip 1.5 cm

\begin{center}
{Ahmad~Shariati{\footnote {shariati@mailaps.org}}\\
Mohammad~Khorrami{\footnote {mamwad@mailaps.org}}\\
Amir~H.~Fatollahi{\footnote {ahfatol@gmail.com}}}
\vskip 10 mm \textit{ Department of Physics, Alzahra University,
Tehran 1993891167, Iran }
\end{center}

\vskip 1.5 cm

\begin{abstract}
\noindent The possibility is examined that a model based on space
noncommutativity of linear type can explain why photons from distant
sources with multi-TeV energies can reach earth. In particular within
a model in which space coordinates satisfy the algebra of SU(2) Lie
group, it is shown that there is the possibility that the pair production
through the reaction of CMB and energetic photons would be forbidden kinematically.
\end{abstract}
\end{titlepage}

\section{Introduction}
There have been arguments supporting the idea
that the ordinary picture of spacetime breaks down when is
measured in intervals comparable with the Planck length and time.
In particular, in an ultra-large momentum transfer experiment a
black-hole may be formed, and as long as it lives before its rapid
evaporation, an outer observer experiences limits on information
transfer from the volume element comparable in size with the
horizon \cite{doplicher}. These kinds of reasoning may lead one to
believe in some kinds of space-space and space-time uncertainty
relations \cite{doplicher}. As uncertainty relations usually point
to noncommutative objects, it is reasonable to consider various
versions of noncommutative spacetime theories. Some important
examples of these kinds of spacetimes are\\
1) the canonical one:
\begin{align}\label{zz01}
[\widehat{x}^{\,\mu},\widehat{x}^{\,\nu}]=\ir\,\theta^{\mu\nu}\,\mathbf{1},
\end{align}
with $\theta^{\mu\nu}$ an antisymmetric constant tensor,\\
2) the $\kappa$-Poinca\'{e} spacetime
\begin{align}\label{zz02}
[\widehat x_a,\widehat t\;]=\frac{\ir}{\kappa}\,\widehat x_a,
\quad[\widehat x_a,\widehat x_b]=0,
\end{align}
where $\kappa$ is a constant, and \\
3) the Lie algebra type
\begin{equation}\label{zz03}
[\widehat x_a,\widehat x_b]=f^c{}_{a\, b}\,\widehat x_c,
\quad[\widehat x_a,\widehat t\;]=0,
\end{equation}
where $f^c{}_{a\,b}$'s are structure constants of a Lie algebra
\cite{kappa,wess,chai,9908142,reviewnc}. In the two latter cases,
$a, b$ and $c$ refer to spatial directions. Recent developments in
understanding the dynamics of D-branes of string theory, have
renewed interest for studying field theories on noncommutative
spacetimes. In particular, the longitudinal directions of D-branes
in the presence of constant B-field background appear to be
noncommutative, as seen by the ends of open strings
\cite{9908142,reviewnc}.

There have been a large number of works devoted to study various
phenomenological implications of noncommutative picture of
spacetime \cite{nc-pheno}. In particular, the two latter
examples mentioned above cause the energy-momentum dispersion
relations as well as the momentum conservation laws differ from
those in the ordinary spacetime \cite{kappa,0612013}. These
different aspects would open a possibility by which one might try
to explain some paradoxical behaviors that have already been
reported in the spectrum of ultrarelativistic particles reaching
the earth from distant sources, one of them is the so-called
``TeV-photon paradox". As photons from very distant sources should
pass through the photon-bath of the $2.7$~K Cosmic Microwave
Background (CMB) radiation, it is expected that photons with
energies of several TeV or higher have very little chance to reach
earth, as they should react with the low energy photons of CMB to
create electron-positron pairs through the reaction
$\gamma\,\gamma\to e^-e^+$. A large number of such multi-TeV,
however, have been reported to reach the earth (see e.g.
\cite{obs-tev}), implying that the universe is perhaps more
transparent to TeV-photons than it should be, the so-called
TeV-photon paradox.

Among other things, there have been attempts aiming to understand
a new physics from the unexpected behavior of energetic particles.
Models based on violation of Lorentz invariance \cite{lorentz},
expectations from quantum gravity effects \cite{qu-gr}, large
extra dimensions \cite{led}, and noncommutativity of
$\kappa$-Poincar\'{e} type \cite{k-tev,k-tev1} are examples. A common
thing in a majority of the above-mentioned approaches is that
perhaps a deformed energy-momentum dispersion relation,
or a deformation of conservation laws, could explain
why such kinds of photons or other ultrahigh-energy particles can
reach the earth. In particular, it could be that these deformations
{\em kinematicaly} forbid such pair creations, or at least push
them to higher energies. The aim of the present paper is to
investigate the effect of a particular noncommutativity of space,
namely a noncommutativity of Lie-algebra type with the Lie algebra
SU(2) \cite{0612013}, on such pair creations. It is shown that
the introduction of such noncommutativity does indeed push the
threshold of pair production to higher energies and could even
forbid it at all.

The scheme of the rest of this paper is the following. In section
2, a brief review is given on kinematical relations as well as
conservation laws in a theory on noncommutative spaces of SU(2)
algebra type. In section 3 the analysis is presented to find the
threshold condition in which the pair production would occur,
based on one can set the parameters to prevent the reaction.
Section 4 is devoted to the concluding remarks.

\section{Conservation laws and dispersion relations}
In \cite{0612013} a model was introduced based on noncommutativity
of Lie algebra type in which, as will be seen in more detail, the
momentum conservation law as well as the energy-momentum
dispersion relation are different from those in the ordinary
space. In particular, a model is investigated in a 3+1 dimensional
space-time the dimensionless spatial position operators of which
are generators of a regular representation of the SU(2) algebra,
that is
\begin{equation}\label{zz04}
[\widehat x_a,\widehat x_b]=\epsilon^c{}_{a\,b}\,\widehat x_c.
\end{equation}
As it was discussed in \cite{0612013}, one can use the group
algebra as the analogue of functions defined on ordinary space,
with group elements $U=\exp(\ell\,k^a\,\widehat x_a)$ as the
analogues of $\exp(\ir\,\mathbf{k}\cdot\mathbf{x})$, which are a
basis for the functions defined on the space. In both cases
$\mathbf{k}$ is an ordinary vector with
$\mathbf{k}=(k^1,k^2,k^3)$. That is the components of $\mathbf{k}$
are commuting numbers. In the case of noncommutative space, $\ell$
is a length parameter, and the vector $\mathbf{k}$ is restricted to a
ball of radius $(2\,\pi/\ell)$, with all points of the boundary
identified to a single point. The manifold of $\mathbf{k}$ is in
fact a 3-sphere. $\mathbf{k}$ can be thought of as the momentum of
a particle. While in the ordinary space one has simple
conservation of momentum in the sense that for a collection of
incoming particles there is a relation
\begin{equation}\label{zz05}
\sum_i\mathbf{k}_i=0,
\end{equation}
in the case of noncommutative space one has (as was shown in
\cite{0612013})
\begin{equation}\label{zz06}
U_1\,U_2\,\cdots=\id,
\end{equation}
or similar equations in which the order of $U_i$'s has been
changed. In particular, if two particles with momenta
$\mathbf{k}_1$ and $\mathbf{k}_2$ collide with each other, the
{\em momentum} of the system is
$\boldsymbol{\gamma}(\mathbf{k}_1,\mathbf{k}_2)$ or
$\boldsymbol{\gamma}(\mathbf{k}_2,\mathbf{k}_1)$ with
\begin{equation}\label{zz07}
\exp[\ell\,\gamma^a(\mathbf{k},\mathbf{k}')\,\widehat
x_a]:=\exp(\ell\,k^a\,\widehat x_a)\,\exp(\ell\,k'^a\,\widehat
x_a).
\end{equation}
The explicit form of $\boldsymbol{\gamma}$ is obtained from
\begin{align}\label{zz08}
\cos\left(\frac{\ell\,\gamma}{2}\right)=&~
\cos\left(\frac{\ell\,k}{2}\right)\,
\cos\left(\frac{\ell\,k'}{2}\right)-
\hat{\mathbf{k}}\cdot\hat{\mathbf{k}'}\,
\sin\left(\frac{\ell\,k}{2}\right)\,
\sin\left(\frac{\ell\,k'}{2}\right),\cr
\hat{\boldsymbol{\gamma}}\,\sin\left(\frac{\ell\,\gamma}{2}\right)=&
~\hat{\mathbf{k}}\times\hat{\mathbf{k}'}\,
\sin\left(\frac{\ell\,k}{2}\right)\,
\sin\left(\frac{\ell\,k'}{2}\right)\cr &+
\hat{\mathbf{k}}\,\sin\left(\frac{\ell\,k}{2}\right)\,
\cos\left(\frac{\ell\,k'}{2}\right)+
\hat{\mathbf{k}'}\,\sin\left(\frac{\ell\,k'}{2}\right)\,
\cos\left(\frac{\ell\,k}{2}\right).
\end{align}
It is easy to see that in the limit $\ell\to 0$,
$\boldsymbol{\gamma}$ tends to $\mathbf{k}+\mathbf{k'}$, as
expected. It is emphasized, as the time direction is not involved
in the noncommutativity relation (\ref{zz03}), the energy
conservation law is just like in ordinary space. As a final
remark, the energy-momentum dispersion relation would come of the
form \cite{0612013}
\begin{equation}\label{zz09}
E=\sqrt{m^2+\left(\frac{4}{\ell}\right)^2\,
\sin^2\left(\frac{k\,\ell}{4}\right)},
\end{equation}
with $m$ as the mass of the particle. Again it is seen that in the
limit $\ell\to 0$ one has $E=\sqrt{m^2+k^2}$, just like the case
of the ordinary space.
\section{Pair production from two colliding photons}
Consider the pair-production reaction $\gamma\,\gamma\to e^-e^+$.
Of the two incoming photons, one is a CMB photon (particle 0), and
the other an energetic one (particle 1). The two outgoing
particles (particles 2 and 3) are of mass $m$. The conservation
laws are
\begin{align}\label{zz10}
E_2+E_3 &= E_0+E_1, \\ \label{zz11} U_2\,U_3 &= U_0\,U_1.
\end{align}
A note is in order concerning the second equality. This is in fact
only one possibility among six possibilities. These possibilities
arise from the fact that in (\ref{zz06}), one can change the order
of $U_i$'s. Instead of changing the order of $U_i$'s, however, one
can use properly-similarity-transformes $U_i$'s with the same
order. So, reminding that in our case at least kinematically the
roles by particles 2 and 3 can be exchanged, one could write
\begin{equation}\label{zz12}
U'_2\,U'_3 = U'_0\,U'_1
\end{equation}
instead of (\ref{zz11}), where $U'_i$'s are obtained from certain
similarity transformations acting on $U_i$'s. One notes that a
similarity transformation on $U(\mathbf{k})$ does not change the
length of $\mathbf{k}$, and hence leaves the dispersion relation
(\ref{zz09}) intact.

It is convenient to move to new variables defined by
\begin{align}\label{zz13}
\epsilon:=&\frac{E}{m},\\ \label{zz14}
u:=&\sin\left(\frac{k\,\ell}{4}\right),\\ \label{zz15}
\mu:=&\frac{m\,\ell}{4}.
\end{align}
One then has for massive particles (2 and 3)
\begin{equation}\label{zz16}
u=\mu\,\sqrt{\epsilon^2-1}.
\end{equation}
where $\epsilon \geq 1$, and for massless particles (0 and 1)
\begin{equation}\label{zz17}
u=\mu\,\epsilon.
\end{equation}
So the energy conservation reads
\begin{equation}\label{zz18}
\epsilon_2+\epsilon_3=\epsilon_0+\epsilon_1.
\end{equation}
In (\ref{zz12}), LH and RH, as the lengths of the momenta
corresponding to the left-hand side and the right-hand side,
respectively, are
\begin{align}\label{zz19}
\mathrm{LH}:=&
\cos\left(\frac{k_2\,\ell}{2}\right)\,\cos\left(\frac{k_3\,\ell}{2}\right)-
\hat{\mathbf{k}'}_2\cdot\hat{\mathbf{k}'}_3\,
\sin\left(\frac{k_2\,\ell}{2}\right)\,\sin\left(\frac{k_3\,\ell}{2}\right),\\
\label{zz20} \mathrm{RH}:=&
\cos\left(\frac{k_0\,\ell}{2}\right)\,\cos\left(\frac{k_1\,\ell}{2}\right)-
\hat{\mathbf{k}'}_0\cdot\hat{\mathbf{k}'}_1\,
\sin\left(\frac{k_0\,\ell}{2}\right)\,\sin\left(\frac{k_1\,\ell}{2}\right),
\end{align}
with
\begin{equation}\label{zz21}
0\leq\frac{k_i\,\ell}{2}\leq \pi.
\end{equation}
One has
\begin{align}\label{zz22}
\mathrm{LH}=\,&\mathrm{RH}.
\end{align}
The aim is to find a threshold for the energy of particle 1, in
order that the reaction occurs. In the case of the ordinary space,
at this the outgoing particles have equal velocities. Here things
look more complicated. Using (\ref{zz19}), (\ref{zz20}), and the
condition (\ref{zz21}), one finds
\begin{align}\label{zz23}
\mathrm{LH}\geq&
\cos\left(\frac{k_2\,\ell}{2}\right)\,\cos\left(\frac{k_3\,\ell}{2}\right)-
\sin\left(\frac{k_2\,\ell}{2}\right)\,\sin\left(\frac{k_3\,\ell}{2}\right),\\
\label{zz24} \mathrm{RH}\leq&
\cos\left(\frac{k_0\,\ell}{2}\right)\,\cos\left(\frac{k_1\,\ell}{2}\right)+
\sin\left(\frac{k_0\,\ell}{2}\right)\,\sin\left(\frac{k_1\,\ell}{2}\right).
\end{align}
At threshold the right-hand sides equal each other, resulting in
\begin{equation}\label{zz25}
\sin\left[\frac{(k_2+k_3)\,\ell}{4}\right]=
\sin\left[\frac{(k_1-k_0)\,\ell}{4}\right].
\end{equation}
One may ask about an altered choice in place of inequalities
(\ref{zz23}) and (\ref{zz24}), in which the role of LH and RH are
exchanged. That choice would give a more sever constraint on
$k_1$. The reason is that if (\ref{zz25}) is satisfied with $k_0$,
$k_1$, and some $k_2$ and $k_3$, then to fulfil the conservation
of momentum with the angle between $\mathbf{k}_0$ and
$\mathbf{k}_1$ less than $\pi$, $k_2$ and $k_3$ should be
increased, meaning that the energy of the outgoing particles
should increase while that of the incoming particles is kept
constant. This violates the energy conservation. The threshold
condition (\ref{zz25}) is in fact similar to its analogue in the
case of the ordinary space, where the collision is head-on and the
outgoing particles move in the same direction.

The condition (\ref{zz25}) can be written like
\begin{equation}\label{zz26}
u_2\,\sqrt{1-u_3^2}+u_3\,\sqrt{1-u_2^2}=
u_1\,\sqrt{1-u_0^2}-u_0\,\sqrt{1-u_1^2}.
\end{equation}
Defining
\begin{equation}\label{zz27}
A:=\epsilon_1\,\sqrt{1-\mu^2\,\epsilon_0^2}-
\epsilon_0\,\sqrt{1-\mu^2\,\epsilon_1^2},
\end{equation}
for which, by (\ref{zz26}), one has
\begin{equation}\label{zz28}
\sqrt{(\epsilon_2^2-1)\,[1-\mu^2\,(\epsilon_3^2-1)]}+
\sqrt{(\epsilon_3^2-1)\,[1-\mu^2\,(\epsilon_2^2-1)]}=A,
\end{equation}
and further defining
\begin{align}\label{zz29}
a:=&\,\frac{\epsilon_0+\epsilon_1}{2}=\frac{\epsilon_2+\epsilon_3}{2}\geq 1,\\ \label{zz30}
x:=&\,\epsilon_2-a = a - \epsilon_3,\\
\label{zz31} z:=&\,x^2,
\end{align}
it is seen that the condition (\ref{zz25}) can be written like
\begin{equation}\label{zz32}
f(z)=0,
\end{equation}
where
\begin{equation}\label{zz33}
f(z):=f_0+f_1\,z+f_2\,z^2,
\end{equation}
with
\begin{align}\label{zz34}
f_0:=&\, 4\,A^2\,(a^2-1)\,[1-\mu^2\,(a^2-1)]-A^4,\\ \label{zz35}
f_1:=&\, 4\,(A^2-4\,a^2+2\,\mu^2\,A^2+2\,\mu^2\,A^2\,a^2),\\
\label{zz36} f_2:=&\, -4\,\mu^2\,A^2.
\end{align}
We remind by (\ref{zz16}) $\epsilon_2,\epsilon_3 \geq 1$, leading
to the condition $-(a-1)\leq x \leq a-1$. So only those roots of
(\ref{zz32}) are acceptable that satisfy
\begin{equation}\label{zz37}
0\leq z\leq (a-1)^2.
\end{equation}
So the aim is to find values of $\mu$ and $\epsilon_1$ so that
(\ref{zz32}) and (\ref{zz37}) are satisfied.

Using the known values of the electron mass $m$ and the energy of
a typical CMB photon, one has
\begin{equation}\label{zz38}
\epsilon_0\sim 10^{-9}.
\end{equation}
So, reminding (\ref{zz18}), and $\epsilon_2,\epsilon_3 \geq 1$ by
(\ref{zz16}), we have
\begin{equation}\label{zz39}
\epsilon_1 > 1.
\end{equation}
$\tilde E_1$ (the threshold energy in the ordinary space)
satisfies
\begin{equation}\label{zz40}
\tilde E_1\,E_0=m^2,
\end{equation}
or
\begin{equation}\label{zz41}
\mu=\frac{\tilde E_1\,\ell}{4}\,\epsilon_0.
\end{equation}
From (\ref{zz09}), however, it is seen that for massless particle~1
\begin{equation}\label{zz42}
E_1\leq \frac{4}{\ell}.
\end{equation}
Assuming photons of the energy $\tilde E_1$ have in fact been observed,
one has
\begin{equation}\label{zz43}
\tilde E_1\leq \frac{4}{\ell},
\end{equation}
resulting in
\begin{equation}\label{zz44}
\mu\leq\epsilon_0.
\end{equation}
Using these, one can approximate $A$ and $a$ to get
\begin{equation}\label{zz45}
A\simeq\epsilon_1-C\,\epsilon_0,
\end{equation}
where
\begin{equation}\label{zz46}
C:=\sqrt{1-u_1^2},
\end{equation}
and
\begin{align}\label{zz47}
\frac{f_0}{A^2}=&~2\,\epsilon_1\,\epsilon_0\,(1+C)-4-\frac{\mu^2}{4}\,
(\epsilon_1^4+4\,\epsilon_1^3\,\epsilon_0-8\,\epsilon_1^2),\\
\label{zz48}
\frac{f_1}{A^2}=&~4\,\left[-2\,\frac{\epsilon_0}{\epsilon_1}\,(1+C)
+2\,\mu^2+\frac{\mu^2\,\epsilon_1^2}{2}+\mu^2\,\epsilon_1\,\epsilon_0\right],\\
\label{zz49} \frac{f_2}{A^2}=&-4\,\mu^2.
\end{align}
Also the discriminator of (\ref{zz32}) becomes
\begin{equation}\label{zz50}
\frac{\Delta'}{16\,A^4}=\mu^4\,\epsilon_1^2-\mu^2
+2\,(1+C)\left(\frac{\epsilon_0}{\epsilon_1}\right)^2.
\end{equation}
If $\epsilon_1\sim 1$, then $f_0$, $f_1$, and $f_2$ are all
negative, showing that for nonnegative $z$ the value of $f(z)$ is
negative. Hence
\begin{equation}\label{zz51}
\epsilon_1\gg 1.
\end{equation}
Using these,
\begin{align}\label{zz52}
\frac{f_0}{A^2}=&\,2\,\epsilon_1\,\epsilon_0\,(1+C)-4-
\frac{\mu^2\,\epsilon_1^4}{4},\\ \label{zz53}
\frac{f_1}{A^2}=&\,4\,\left[-2\,\frac{\epsilon_0}{\epsilon_1}\,(1+C)
+\frac{\mu^2\,\epsilon_1^2}{2}\right],\\ \label{zz54}
\frac{f_2}{A^2}=&-4\,\mu^2,\\ \label{zz55}
\frac{\Delta'}{16\,A^4}=&\,\mu^4\,\epsilon_1^2-\mu^2
+2\,(1+C)\,\left(\frac{\epsilon_0}{\epsilon_1}\right)^2.
\end{align}
The condition that $\Delta'$ be nonnegative is then
\begin{equation}\label{zz56}
u_1^2\,(1-u_1^2)\leq 2\,(1+C)\,\epsilon_0^2.
\end{equation}
Reminding (\ref{zz38}), this shows that $u_1$ is either
close to zero or close to one. So, there are two cases:\\
\textbf{Case I} $(u_1\simeq 0)$:
\begin{equation}\label{zz57}
C\simeq 1,
\end{equation}
yielding
\begin{equation}\label{zz58}
0\leq u_1\lesssim 2\,\epsilon_0,
\end{equation}
by (\ref{zz56}). As $f_2$ is negative, in order that there exists a positive
solution for (\ref{zz32}), at least one of $f_0$ or $f_1$ should
be positive. The condition that $f_1$ be positive, is
\begin{equation}\label{zz59}
u_1^3\gtrsim 8\,\mu\,\epsilon_0.
\end{equation}
This together with (\ref{zz58}) gives a necessary condition that
$f_1$ be positive:
\begin{equation}\label{zz60}
\frac{\mu}{\epsilon_0^2}\leq 1.
\end{equation}
The condition that $f_0$ be positive is that
\begin{equation}\label{zz61}
4\,(\epsilon_0\,\epsilon_1)-4-\frac{1}{4}\,
\left(\frac{\mu}{\epsilon_0^2}\right)^2\,(\epsilon_0\,\epsilon_1)^4\geq 0,
\end{equation}
which gives
\begin{equation}\label{zz62}
\frac{\mu}{\epsilon_0^2}\leq
4\,\sqrt{\frac{(\epsilon_0\,\epsilon_1)-1}{(\epsilon_0\,\epsilon_1)^4}}.
\end{equation}
So, by the maximum of right-hand-side of (\ref{zz62}), we have
\begin{equation}\label{zz63}
\frac{\mu}{\epsilon_0^2}\leq\frac{\sqrt{27}}{4}.
\end{equation}
If both (\ref{zz60}) and (\ref{zz63}) are violated, then there is
no positive solution for (\ref{zz32}).
\\
\textbf{Case II} $(u_1\simeq 1)$:
\begin{equation}\label{zz64}
C\simeq 0,
\end{equation}
yielding
\begin{equation}\label{zz65}
1-\epsilon_0^2\lesssim u_1\leq 1,
\end{equation}
by (\ref{zz56}). Defining
\begin{align}\label{zz66}
w:=&\,z-(a-1)^2,\\ \label{zz67} \tilde f(w):=&\, f(z),
\end{align}
where
\begin{equation}\label{zz68}
\tilde f(w)=\tilde f_0+\tilde f_1\,w+\tilde f_2\,w^2,
\end{equation}
one arrives at
\begin{align}\label{zz69}
\frac{\tilde f_0}{A^2}=&-8,\\
\label{zz70} \frac{\tilde f_1}{A^2}=& ~8\,\mu^2\,\epsilon_1,
\end{align}
showing that $\tilde f_0$ is negative and $\tilde f_1$ is
positive. This shows that no nonpositive $w$ exists so that
$\tilde f(w)$ vanishes. This shows that in this case no solution
exists.

To summarize, the threshold occurs at $u_3=u_2$, and no reaction
is possible if
\begin{equation}\label{zz71}
\mu\geq\frac{\sqrt{27}}{4}\,\epsilon_0^2,
\end{equation}
or equivalently
\begin{equation}\label{zz72}
\frac{4}{\ell}\leq\frac{4}{\sqrt{27}}\,\frac{m^3}{E_0^2}.
\end{equation}
By inserting the available values for $E_0$ and $m$ we see that for
$\ell \sim 10^{-11}~\textrm{TeV}^{-1}$ or higher, the reaction
would be forbidden. The value corresponds to highest energy
up to $4\times 10^{11}$~TeV, which in principle can reach and be
detected safely.

\section{Concluding remarks}
It was seen that introducing SU(2)-fuzziness in the space
coordinates shifts the energy threshold for the pair-production
reaction, in which an energetic photon reacts with a photon
of the cosmic microwave background. Besides, if the energy
corresponding to the fuzziness length is smaller than a certain
value, this reaction would be forbidden. This is essentially due
to the fact that introducing a Lie-group fuzziness in the space
coordinates, makes the momenta space compact iff the group is
compact, hence introducing a cutoff for the momenta (and energy).
It would be useful to compare the situation here with the case of
fuzziness of $\kappa$-Poincar\'{e} type \cite{k-tev,k-tev1}. As
discussed in \cite{k-tev1} based on the key result Eq.~(15), the
threshold value is not changed in $\kappa$-Poincar\'{e} case, at
least up to the leading order corrections introduced by $\kappa$
(see item (ii) after Eq.~(15)). Instead, their result shows that
for pair production above threshold the process is possible only
by an energy value higher than the value corresponding to the
classical limit, $\kappa\to \infty$ \cite{k-tev1}. In this case the
momentum space is not compact but reduced relative to the
classical case. It has been there argued that an explanation for
the exotic behavior of TeV-photons might be possible
\cite{k-tev1}. A notable difference between that work and the
present work is that here the momentum space is compact, and this
could even make the pair-production reaction impossible.

\vspace{0.5cm}
\textbf{Acknowledgement}:  This work was partially supported by
the research council of the Alzahra University.

\end{document}